# Band structure of the epitaxial Fe/MgO/GaAs(001) tunnel junction studied by X-ray and ultraviolet photoelectron spectroscopy

Y. Lu, J. C. Le Breton, P. Turban,[a]  B. Lépine, P. Schieffer, and G. Jézéquel

*Equipe de Physique des Surfaces et Interfaces, UMR 6627, CNRS-Université de Rennes 1, Campus de Beaulieu, Bât 11C, 35042 Rennes cedex, France*

**Abstract**

The electronic band structure in the epitaxial Fe/MgO/GaAs(001) tunnel junction has been studied by X-ray and ultraviolet photoelectron spectroscopy measurements. The Schottky barrier height (SBH) of Fe on MgO/GaAs heterostructure is determined to be 3.3±0.1eV, which sets the Fe Fermi level about 0.3eV above the GaAs valence band maximum. This SBH is also exactly the same as that measured from Fe on MgO monocrystal. After Fe deposition, no band bending change is observed in MgO and GaAs underlayers. On the contrary, Au and Al deposition lead to clear variation of the band bending in both MgO and GaAs layers. This effect is analyzed as a fingerprint of defects states at the MgO/GaAs interface.

*a) Corresponding author, e-mail: pascal.turban@univ-rennes1.fr*

Significant research effort has been devoted to study the spin injection from a transition metal into semiconductors, due to the large panel of potential applications in the field of sensors, logical devices, or memories.[1] It is now appreciated that the use of a tunnel barrier between the ferromagnetic metallic electrode and the semiconductor can lead to high efficiency spin injection.[2,3] Recently, a CoFe/MgO(001) tunnel injector has been used for room-temperature highly spin-polarized current injection into GaAs.[4] Due to its relative ease of preparation and thermal stability,[4,5] the CoFe/MgO/GaAs(001) solid state spin injector appears as a very promising candidate for future spintronic applications. To completely understand tunnel spin injection in the Fe/MgO/GaAs system, the electronic band-structure of this model system should be first precisely known. In this letter a combined X-ray and ultraviolet photoelectron spectroscopy (XPS and UPS) study is used to determine the band-structure lineup of the Fe/MgO/GaAs(001) heterostructure. The Schottky barrier height (SBH) at the Fe/MgO interface and the band bending in the MgO barrier are especially measured.

The epitaxial Fe/MgO/GaAs(001) samples were grown by molecular beam epitaxy (MBE). Details about the preparation of the MgO/GaAs(001) samples can be found elsewhere.[6] Briefly, 3nm thick MgO(001) layers were grown on Si n-doped ($4\times10^{16}$cm$^{-3}$) GaAs As(2×4) surface at room temperature (RT) by evaporation of high purity MgO powder. After MgO growth, the samples were annealed under ultra-high vacuum at 600°C for 20min. Finally, a 2nm thick Fe(001) layer was epitaxially grown at RT on the annealed MgO/GaAs heterostructure. MgKα (1253.6eV) was used as the X-ray source and HeI (21.2eV) resonance line provided the UPS source for photoemission experiments. The total energy resolutions were about 800meV and 100meV for XPS and UPS, respectively.

Fig. 1 schematically shows the energy level diagram of the Fe/MgO/GaAs system. In our previous work, we precisely determined by XPS a valence-band offset (VBO) $\Delta E_V = 4.2 \pm 0.1 eV$ and a conduction-band offset (CBO) $\Delta E_C = 2.2 \pm 0.1 eV$ at the MgO/GaAs(001) interface.[6] As XPS measurements are slightly averaged in depth, band bending in the MgO layer could induce a systematic error in our measurements. To check this point, we have grown MgO with different thicknesses (2nm, 3nm and 4nm) on GaAs. It was found that O2s and Ga3d core level positions of these samples do not change (variation smaller than $\pm 0.05 eV$), which proves that the MgO layer is almost in flat band conditions.

In this paper, the Schottky barrier height is measured on the Fe/MgO/GaAs(001) metal/insulator/semiconductor (MIS) sample. To determine the SBH $\Phi_{BS}$ of Fe on MgO (the Fe Fermi level (FL) position with respect to the MgO conduction band minimum (CBM)), the MgO O2s core level position is used as a reference peak:

$$\Phi_{BS} = E_g^{MgO} + (E_{O2s}^{MgO} - E_{VBM}^{MgO}) - \Delta E_{FC}, \qquad (1)$$

where $E_g^{MgO}$ is MgO band gap, and $\Delta E_{FC} = (E_{O2s}^{MgO} - E_F)$ is the energy difference between the MgO O2s core level and the Fe FL which are measured from the MIS sample. Since the energy difference of O2s to MgO valence band maximum (VBM) $(E_{O2s}^{MgO} - E_{VBM}^{MgO})$ has already been carefully determined to be 18.06eV,[6] the SBH $\Phi_{BS}$ can be deduced from the measurement of $\Delta E_{FC}$.

Fig. 2 shows XPS spectra of O2s, Ga3d core levels and Fe FLs for different samples, respectively 3nm MgO/GaAs before Fe deposition, 3nm MgO/GaAs after 2nm Fe deposition, and 2nm Fe deposited on an UHV-cleaved MgO monocrystal. The Fe thickness (2nm) is thick enough to achieve the percolation of the Fe islands and to observe a bulk-like Fe valence-band character. The Fe FL position is taken at the maximum slope of the rising edge of the primary valence electrons. In addition, the Fe FL position of the sample on MgO monocrystal is aligned to that of the MIS sample to be able to compare their O2s positions. It is interesting to note that their O2s core levels are located exactly at the same position. As a result, the distance $\Delta E_{FC}$ is measured to be 22.55±0.05eV. Substitution of the $E_g^{MgO}$ (7.83eV[7]), ($E_{O2s}^{MgO} - E_{VBM}^{MgO}$) and $\Delta E_{FC}$ values into Eq.(1) results in the SBH of Fe on MgO:

$$\Phi_{BS} = 7.83 + 18.06 - 22.55 \approx 3.3 \pm 0.1 eV \tag{2}$$

The SBH value is found to be the same for Fe on MgO/GaAs as for Fe on MgO monocrystal. Combining with the result of MgO/GaAs valence band offset, it is found that the Fe Fermi level lies about 0.3eV above the GaAs VBM.

In Fig. 2, it is also found that the O2s and Ga3d peak positions nearly do not change before and after Fe deposition. If the 3nm MgO layer grown on GaAs presents a very high density of localized defects, the Fe Fermi level could be pinned by those defect states, so that the O2s position would not change after Fe deposition. To clarify this point, 2nm thick Au and Al layers were grown on the MgO/GaAs heterostructure to compare with the Fe case (Pauling's Electronegativity: Al 1.61, Fe 1.83, Au 2.54). Fig. 3 shows the O2s and Ga3d core levels and metal Fermi level spectra recorded by XPS and UPS HeI respectively. The three metals' FLs are found to be exactly at the same position as the system FL. Compared with the case of Fe, the O2s and Ga3d core levels demonstrate clear shifts: towards lower binding energy after Au deposition and towards higher binding energy after Al deposition. The O2s core level position shifts much more than that of Ga3d, so that the distance between O2s and Ga3d core levels changes after different metals deposition. The clear shift of O2s position after Au and Al deposition indicates that the defects density at the metal/MgO interface is too low to pin the Fermi level, which position shows a clear trend with the metals electronegativities as will be discussed below. In addition, the shift of Ga3d core levels also indicates that the Fermi level is not totally pinned on the defects states at the MgO/GaAs interface.

Fig. 4 presents the band bending diagrams for these three MIS structures deduced from the position of O2s and Ga3d core levels. We assumed that the band offsets at the MgO/GaAs interface are fixed and that the band bending in MgO is linear. From the Ga3d position, the position of MgO VBM at the MgO/GaAs interface can be deduced through the MgO/GaAs VBO. In addition, the MgO VBM position at the metal/MgO interface can be obtained from the O2s position measured in the MIS sample.[8] Finally, in the case of Au, there exists a large band bending in the MgO layer, and the potential variation across the MgO layer is about 0.5eV. After Fe deposition, the MgO layer remains in flat band conditions, while in the case of Al, the band bending in the MgO layer is inverted with a potential drop of about 0.4eV. The SBHs of metals on MgO are also measured (Al: 2.8eV, Fe: 3.3eV, Au: 4.0eV).[8] These values are in a good agreement with SBH measured on metals deposited on MgO monocrystal.[9]

In order to understand the physical mechanisms governing bands alignments in these MIS structures, one needs first to consider the pining of the FL position at the MgO/GaAs interface before metals deposition. With the MgO deposition conditions used in this study, the system FL is pinned 0.27eV above GaAs VBM after MgO growth and annealing. This FL position implies a depleted and slightly inverted zone in GaAs (estimated to $7.5 \times 10^{11}$ positive charges per cm$^2$ in GaAs) and a corresponding opposite negative charge on the MgO side, located on acceptors levels in the oxide bandgap. These acceptors levels may originate from interfacial structural defects due to the highly mismatched epitaxial MgO growth on GaAs, or from reactivity at the MgO/GaAs interface as well. On the other hand, the FL position at the metal/MgO interface is governed by the physics of Schottky barriers formation[10]: the SBHs values increase quasi linearly as expected with the increase of metal electronegativities. An exhaustive discussion of the Schottky barriers formation mechanisms[10,11] on MgO including quantitative analysis of the SBH dependence versus metals electronegativities will be published elsewhere.[9]

Finally, the formation of the complete MIS structure implies that the FLs at the MgO/GaAs and metal/MgO interfaces equalize at thermal equilibrium. This equalization is possible thanks to charge transfer in the structure creating the necessary dipoles equilibrating the FLs. In the case of Fe, no charge transfer is needed since the FLs positions at the MgO/GaAs and Fe/MgO interfaces almost match. On the contrary, in the case of Al, electrons are transferred from the metal side towards both the acceptors states in MgO bandgap and GaAs, thus reducing the band bending in the semiconductor. Last, in the case of Au, the charge transfer is inverted and electrons are transferred from both MgO acceptors states and GaAs towards the metallic film, thus increasing the band bending in the semiconductor. Such metal-overlayer-induced charge-transfer has also been observed in metal/$SiO_2$/Si systems.[12] Assuming purely interfacial defects, an optical dielectric constant of 3.18 for MgO and a constant density of the defects states, a classical capacitor model can be used to estimate the acceptors defects states density close the Fermi level at the MgO/GaAs interface. The obtained rough value is $1 \times 10^{13}$ states/$cm^2 \cdot eV$. Such a high defects density should be considerably reduced before future transport applications.

To conclude, we have determined the band structure lineup of Fe/MgO/GaAs(001) heterostructure by XPS and UPS. The SBH of Fe on MgO/GaAs is determined to be 3.3±0.1eV, which sets the Fe Fermi level about 0.3eV above GaAs VBM. The modification of this band lineup by deposition of Al and Au allows us to demonstrate the presence of a large defects states density at the MgO/GaAs(001) interface.

**Acknowledgments**

This work was supported by the National French Program PNANO (MOMES project).

**Figures**

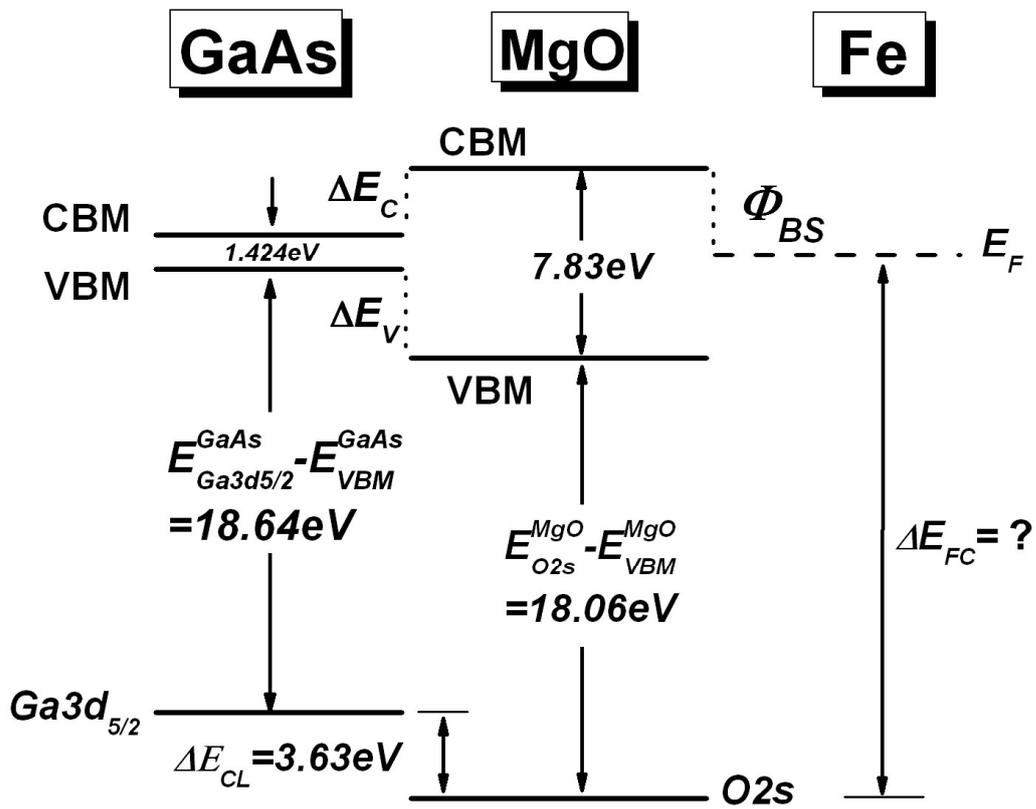

FIG. 1: Schematic energy level diagram of Fe/MgO/GaAs heterostructure.

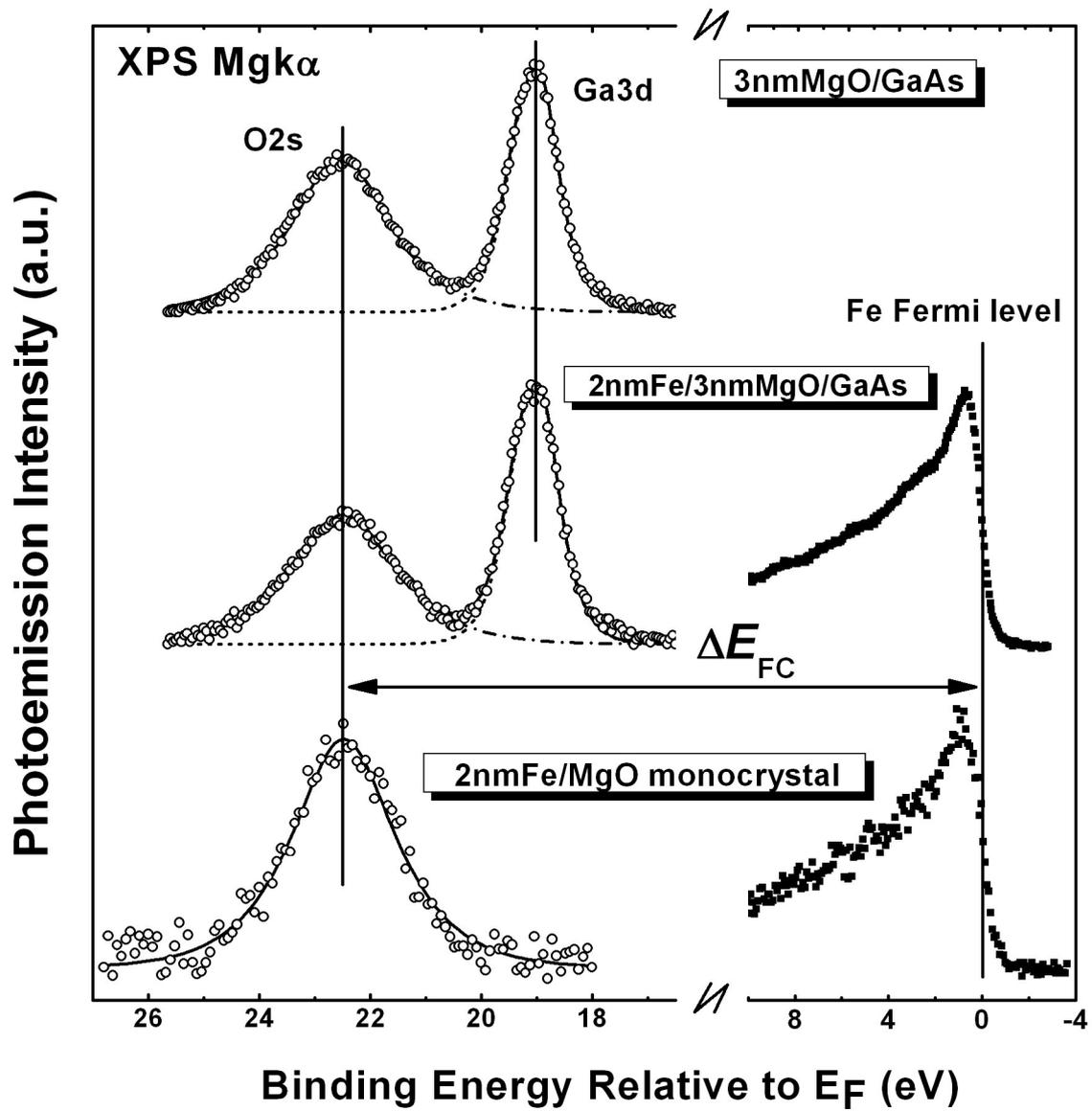

FIG. 2: XPS spectra of O2s, Ga3d core levels and Fe Fermi levels for different stacks: 3nm MgO/GaAs(001), 2nm Fe/3nm MgO/GaAs(001) and 2nm Fe/MgO(001) monocrystal.

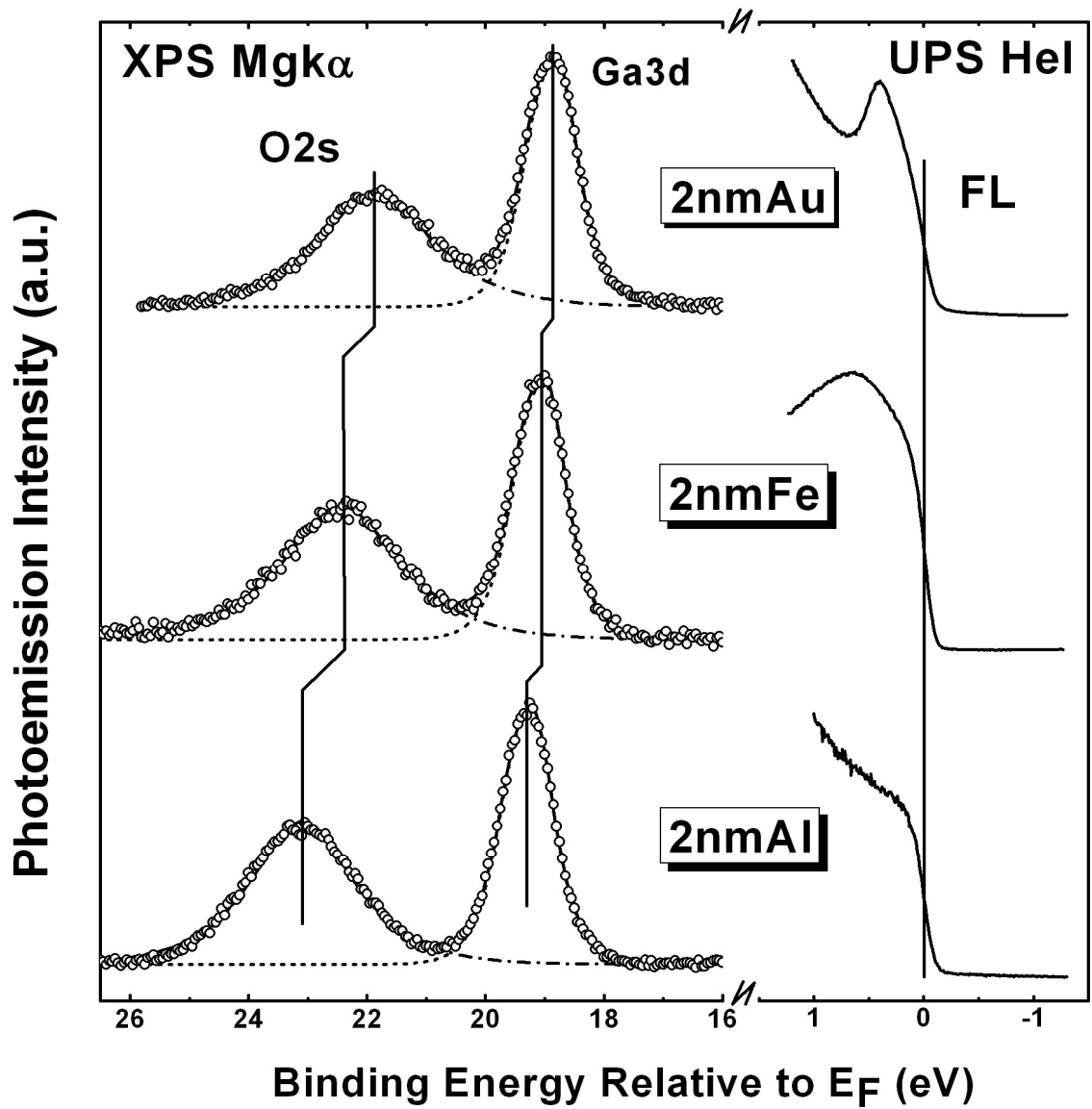

FIG. 3: XPS spectra of O2s and Ga3d core levels for 2nm Au, 2nm Fe and 2nm Al on MgO/GaAs heterostructures. The metal Fermi levels were measured by UPS HeI.

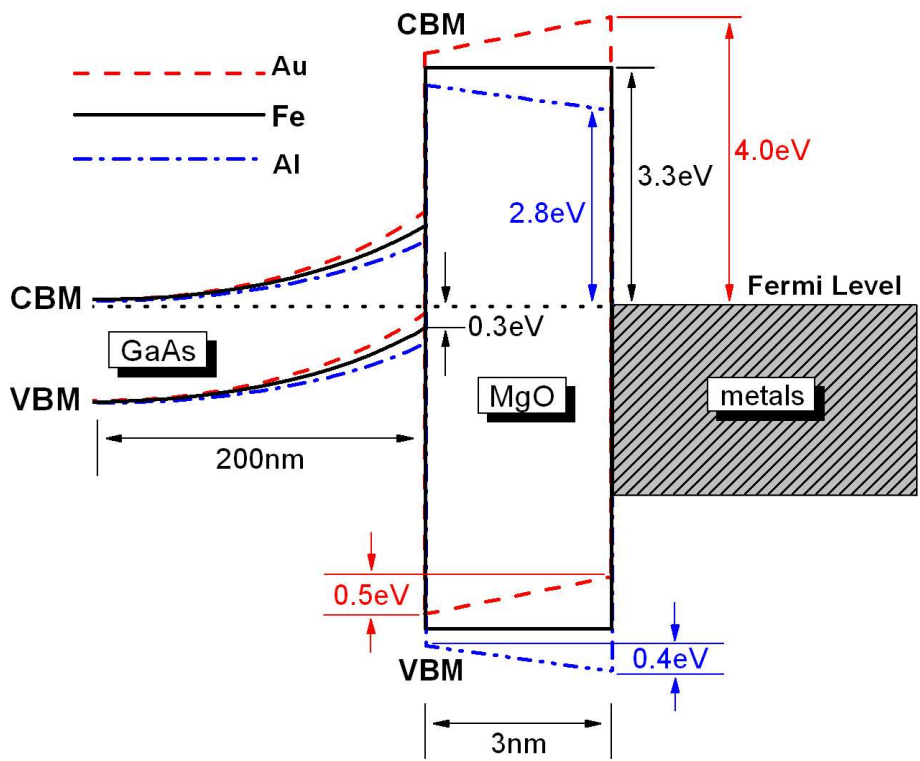

FIG. 4: (Color online) Band bending diagrams for three kinds of MIS structures: Au, Fe and Al on MgO/GaAs(001).